\documentclass{optica-article}
\usepackage{float}
\usepackage{caption}
\usepackage{subcaption} 
\usepackage[abs]{overpic}

\newcommand{\micron}[0]{\mbox{\textmu{m}}}

\journal{opticajournal}  % for journals or Optica Open

\articletype{Research Article}

\begin{document}

\title{Accessing different higher-order modes with nonlinear modal energy transfer under simple realistic tuning of initial conditions}

\author{Julien Dechanxhe, Pascal Kockaert, and Spencer W. Jolly\authormark{*}}

\address{Service OPERA-Photonique, Université libre de Bruxelles (ULB), Brussels, Belgium}

\email{\authormark{*}spencer.jolly@ulb.be} %% email address is required; see note below about the corresponding author designation

% use {asbstract*} to suppress the copyright line. Copyright information will be added in production

\begin{abstract*} 
The initial conditions in multi-mode fibers pumped by ultrashort laser pulses strongly determine the following nonlinear optical interactions. In this work we firstly compare the detailed spatial mode content of simple initial conditions, transverse offset and tilt. We then show how those initial conditions can both be used to achieve a nonlinear modal energy transfer into higher-order spatial modes of a model graded-index fiber, with their own slight differences and advantages. Going beyond purely spatial initial conditions, we introduce nonlinear modal energy transfer results using spatial chirp at the input facet, whereby the different temporal envelopes of the spatial modes allow for tuning the nonlinear modal energy transfer process. Our results open up investigations into higher-dimensional tuning of nonlinear processes in multi-mode fibers using initial conditions.
\end{abstract*}

%%%%%%%%%%%%%%%%%%%%%%%%%%  body  %%%%%%%%%%%%%%%%%%%%%%%%%%
\section{Introduction}
\label{sec:introduction}

Multi-mode fibers, optical fibers that are larger in core diameter and therefore admit more than one and sometimes a large number of propagating modes, have long been imagined as platforms for high-power lasers due the possibility of larger guided profiles. However, small perturbations in the fiber or weak interactions between modes can readily cause a degradation of mode quality, hence making the application to high-power lasers or typical telecommunications devices difficult. The complexities and high spatially frequency content of the linear propagation has however been utilized for enhanced imaging through multi-mode fiber endoscopes~\cite{cizmar12,ploeschner15,caoH23}. More recently, nonlinear propagation and modal interactions have become very active towards rich nonlinear physics and multi-mode photonics devices~\cite{krupa19,wright22-1,wright22-2}.

One of the very interesting results of this recent revival of multi-mode nonlinear optics is that, when modal interactions and input power are significant enough, exchange of energy between the modes of an initially multi-modal beam can lead to a beam self-cleaning (BSC) into the lowest-order mode~\cite{liuZ16,krupa17}. This has spurred an interesting fundamental discussion regarding the nuances of the physical processes~\cite{pourbeyram22,baudin23}, and a number of further demonstrations where tuning the initial conditions can lead to BSC into specific higher-order modes~\cite{deliancourt19-1,deliancourt19-2,chenJiayang22}.

Besides BSC, multi-mode interactions have been shown to tune supercontinuum generation~\cite{wright15-2,wright15-3,eftekhar17}, create interesting multi-mode soliton interactions~\cite{wright15-1,sunY22,sunY24}, and even enable spatial-temporal mode-locking~\cite{wright17,wright20} pushing toward enabled technologies. Beyond typical multi-mode fibers like the graded-index fibers used in most of the cited works, large diameter hollow-core fibers have also been a platform for multi-mode phenomena~\cite{safaei20,piccoli21,brahms22} pushing the frontier of short pulse duration and high pulse energy.

In our work we investigate a process that relies on similar nonlinear inter-modal interactions as beam self-cleaning. We investigate the nonlinear transfer of energy from a standard Gaussian beam while considering tuning simple and physically-relevant initial conditions to have the majority of the energy transfer to even higher-order modes than presented in comparable past work. Specifically, we focus on tilt (propagation direction), offset (central spatial position) and waist (spatial width) of the input Gaussian beam, and their effects on the initial conditions at the fiber facet. We then see how the initial conditions affect nonlinear propagation and transfer of energy into higher-order modes. We focus on experimentally and theoretically simpler initial conditions than those of past experimental work that relied on wavefront shaping~\cite{deliancourt19-2} or simulations that relied on an arbitrary energy distribution~\cite{graini23}. Finally, we discuss using relatively simple spatio-temporal initial conditions, where each coupled mode has a different temporal profile, to further tune the nonlinear energy transfer process.

Our work is purely numerical. The modeling of nonlinear multi-mode fiber propagation is a difficult general problem due to the complexity and/or resolution of the dynamics, and can be done in a number of ways~\cite{poletti08,tani14,bejot19} with different advantages. We use an open-source code~\cite{wright18,github_GMMNLSE} to perform nonlinear propagation simulations that is based on the Generalized Multimode Nonlinear Schrödinger Equation (GMMNLSE). This model relies on modal decomposition. It first computes the spatial overlap between the modes. Then, it propagates each mode individually by considering nonlinear contribution of the other modes with extra terms related to the overlap. The linear part of the model incorporates dispersion up to the fourth order while the nonlinear part includes both instantaneous Kerr nonlinearities and delayed Raman nonlinearities. Importantly, the random linear mode coupling is not considered in this model, which limits the application of our results to short and undisturbed optical fibers. Note that the model is scalar. To account for polarization, we would need to double the number of modes considered, which would considerably increase computation times. Fortunately, this scalar model is valid for describing beam self-cleaning (or nonlinear modal energy transfer), as it has been experimentally demonstrated that this multimode nonlinear process does not depend on the polarization state of the input pulse-beam~\cite{ferraro23}. The code is designed to be easily parallelized, thereby accelerating computation.

Prior to simulating propagation, it is essential to determine the complex initial conditions in each mode, which we discuss in the following section. Following that we describe specific scenarios using different initial conditions to target certain output modes. We present the first demonstration of using a space-time coupling to control such nonlinear interactions, and finally present our conclusions.

\section{Mode Coupling}
\label{sec:mode_coupling}

\begin{figure}[H]
	\centering
	% Première ligne
	\begin{subfigure}{1.0\textwidth}
		\begin{overpic}[width=\textwidth,trim={0cm 5cm 0cm 0cm},clip]{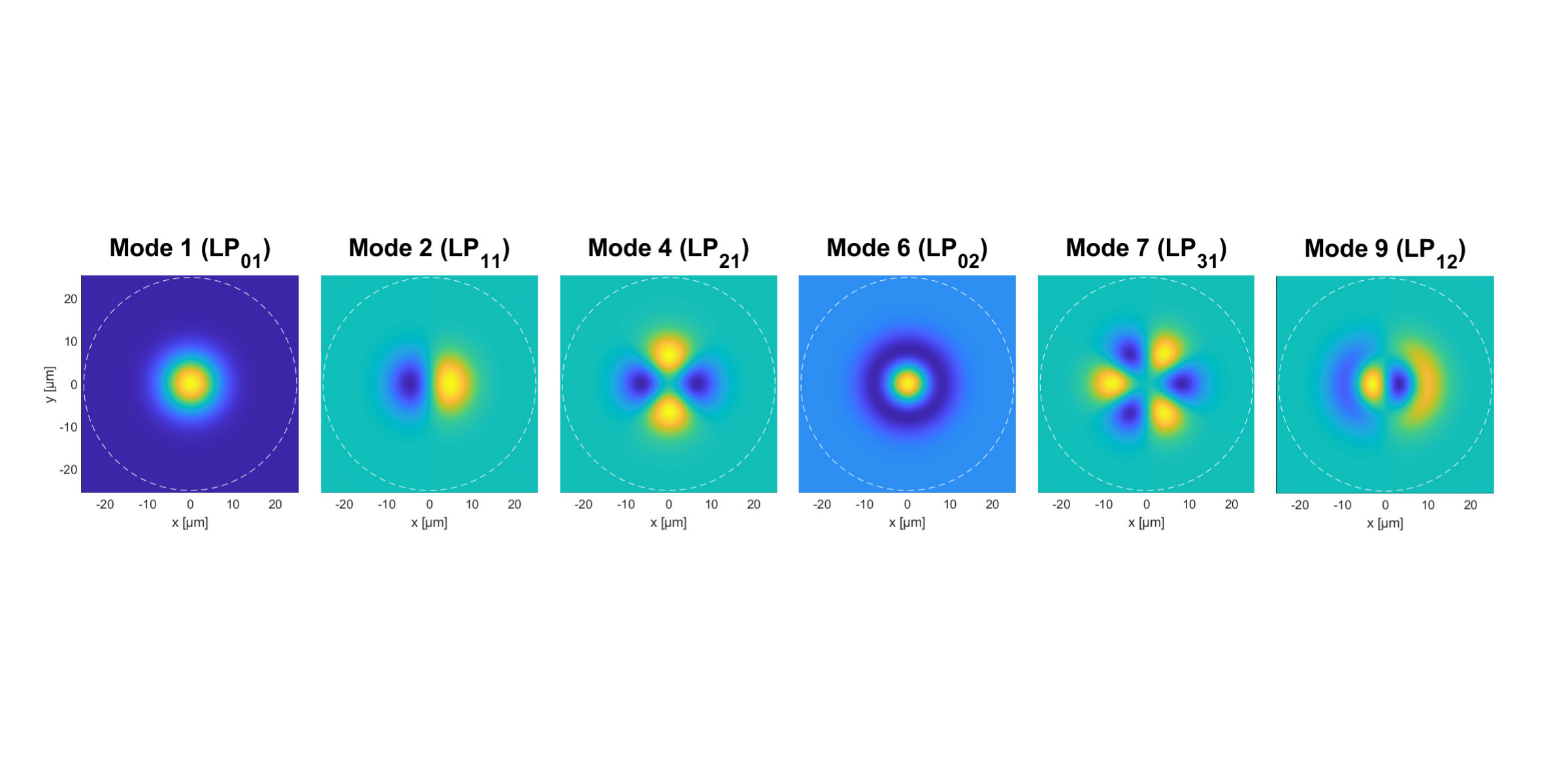}
			\put(10,80){a)}
		\end{overpic}
	\end{subfigure}
	
	% Deuxième ligne
	\vspace{0.5em} % Espacement vertical entre les lignes
	\begin{subfigure}{0.4\textwidth}
		\begin{overpic}[width=1.0\textwidth]{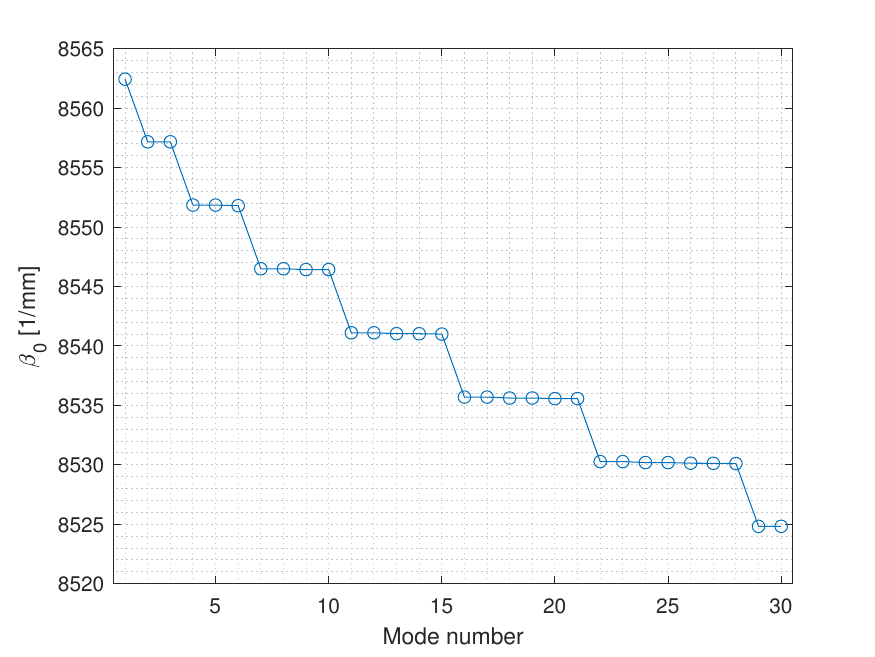}
			\put(-10,100){b)}
		\end{overpic}
	\end{subfigure}
	\hspace{0.75em}
	\raisebox{-0.3cm}{
		\begin{subfigure}{0.4\textwidth}
			\begin{overpic}[width=0.9\textwidth]{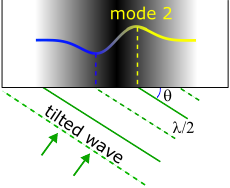}
				\put(-15,110){c)}
			\end{overpic}
		\end{subfigure}
	}
	
	% Troisième ligne
	\vspace{0.1em} % Espacement vertical entre les lignes
	\raisebox{0.3cm}{
		\begin{subfigure}{0.3\textwidth}
			\begin{overpic}[width=0.8\textwidth]{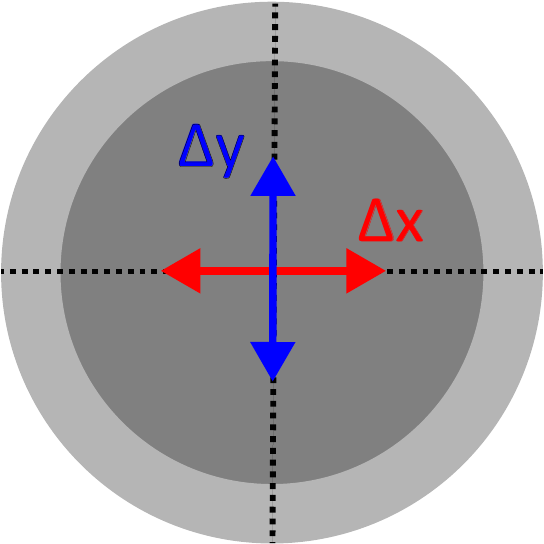}
				\put(0,80){d)}
			\end{overpic}
		\end{subfigure}
	}
	\hspace{0.75em}
	\begin{subfigure}{0.3\textwidth}
		\begin{overpic}[width=1.2\textwidth,trim={2cm 6cm 2.5cm 10cm},clip]{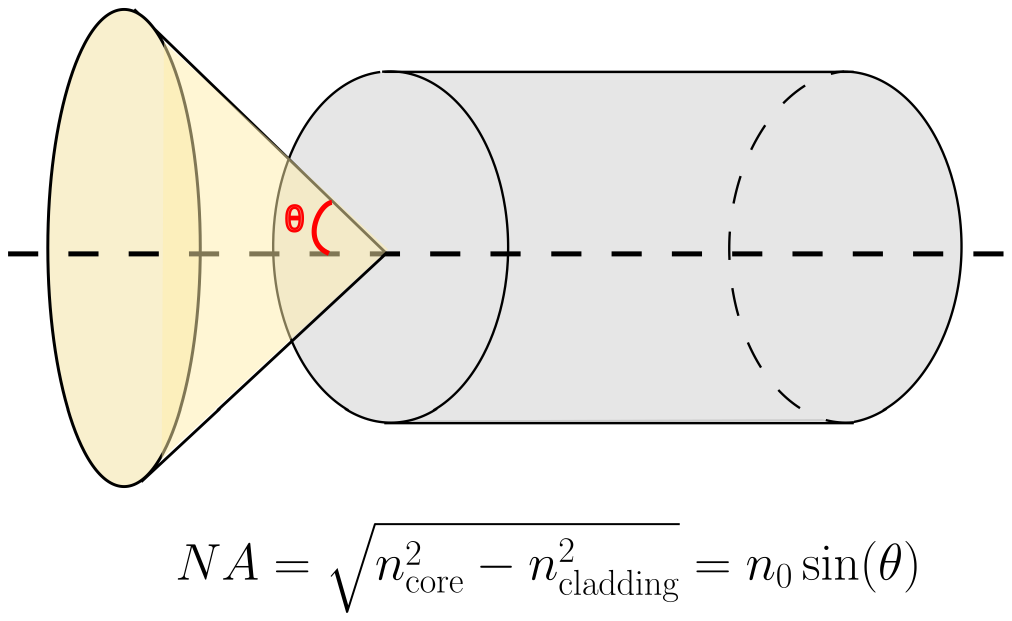}
			\put(0,90){e)}
		\end{overpic}
	\end{subfigure}
	\caption{a) Transverse profiles of modes that are even in the $y$-direction. The equivalent LP notation is also given. The outline of the $50\,\mu m$ diameter fiber core is shown in white dotted line. b) Distribution of propagation constants against mode number. c) Spatial phase matching of a tilted input pulse with mode 2. $\theta \in [0^{\circ},\; 5^{\circ}]$ is the tilt angle. d) Offset in transverse directions ($\Delta x \in [0\,\text{µm}, 15\,\text{µm}]$). e) Numerical aperture ($NA=0.2$) and its associated acceptance cone.}
	\label{CI_Modes}
\end{figure}

% 1st part of the paper : energy distribution into the propagation modes
When a light pulse-beam is injected into a multimode fiber, its energy is distributed among the various propagating (spatial) modes based on the properties of the light beam at the input facet. Those modes for a graded-index fiber of 50\,µm core diameter at 1064\,nm central wavelength are shown in figure~\ref{CI_Modes}.a, where the ordering is according to the decreasing propagation constant in our case rather than the radial or rotational order of the mode. In figure~\ref{CI_Modes}.b, the distribution of propagation constants against mode number is shown for the first 30 modes of the fiber. The ladder-like structure is a typical feature of parabolic-graded index multi-mode fiber and is due to a constant discrepancy between the propagation constants of different non-degenerated groups. The energy distribution in each mode is described by the coupling coefficient~\cite{niuJ07,guangZ18,jolly23-2}:

\begin{equation}
	\eta_i = \frac{\left|\int\int p(x,y,\Delta z)F_i(x,y)dxdy\right|^2}{\int\int \left|p(x,y,\Delta z)\right|^2dxdy\int\int \left|F_i(x,y)\right|^2dxdy},
	\label{eq:coupling_ceoff}
\end{equation}
where $F_i$ represents the transverse profile of the mode, and $p$ is the spatial profile of the injected pulse. This coupling coefficient quantifies the overlap between the beam and the mode of interest, with a unique coefficient for each mode. The method of injecting a light pulse into the fiber has a significant influence on the coupling coefficient, as it alters the overlap between the pulse and the different modes. Key parameters that affect this overlap include the transverse~\cite{eftekhar17} and longitudinal~\cite{ahsan20} offsets, the beam waist (intensity half-width measured at $1/e^2$), and the tilt angle~\cite{deliancourt19-1,brahms22} of the injected beam. In this analysis, a zero longitudinal offset is assumed, although in past work that has also shown to affect nonlinear processes~\cite{ahsan20}. It is important to note that these parameters represent experimental degrees of freedom, i.e. our aim is to work only with initial conditions that are relevant to simple and realistic experimental scenarios.

In the specific case of zero transverse offset and zero tilt angle (see figure~\ref{CI_Modes}), the energy of a Gaussian pulse is not coupled into modes that are antisymmetric in the x- or y-direction, as the overlap integral evaluates to zero. Under these conditions, only modes exhibiting cylindrical symmetry are excited (e.g., modes 1, and 6 according to our numbering, presented in figure~\ref{CI_Modes}.a and~\ref{CI_Modes}.b). When a nonzero transverse offset is introduced (see figure~\ref{CI_Modes}.d), the symmetry is broken, resulting in nonzero coupling coefficients for antisymmetric modes. For instance, if there is a nonzero offset in the x-direction then even modes in the y-direction are excited (see figure~\ref{CI_Modes}.a) while odd modes in y-direction have still zero coupling.

\begin{figure}[H]
	\centering
	\includegraphics[width=1.0\linewidth]{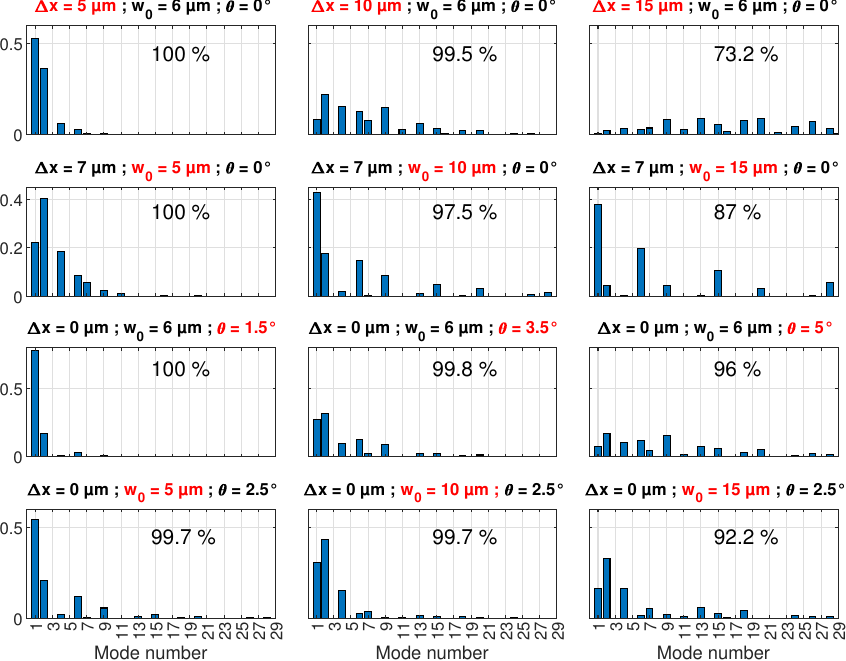}
	\captionsetup{width=1.0\textwidth, justification=raggedright}
	\caption{Energy distribution into the propagation modes of the fiber at its entry. For each panel of the figure a different parameter was modified. The
		modified parameters are (from top to bottom): Offset in x-direction ($\Delta{x}$), waist ($w_0$) at a fixed offset, tilt angle ($\theta$), and again waist but at a fixed tilt angle. The percentage in each panel gives the fraction of the energy that is guided in that case.}
	\label{Graph_Energy_Distribution}
\end{figure}

The first panel of figure~\ref{Graph_Energy_Distribution} illustrates that the energy coupled into the fundamental mode decreases with increasing offset. This occurs because the fundamental mode has a finite size. As the input pulse is injected with a greater offset, its overlap with the fundamental mode diminishes, ultimately resulting in an almost zero-coupling coefficient at an offset of 15\,µm in the x-direction. Since each mode is centered within the fiber and the size of the mode increases with its order (see figure~\ref{CI_Modes}.a), a larger transverse offset leads to the excitation of higher-order modes. Additionally, the total energy transmitted into the fiber decreases with increasing offset. This is mainly because, at a certain offset, there won't be any guided modes that overlap with the portion of the input beam the furthest from the fiber center, i.e. that portion is directly unguided and leaves the fiber.

The second row of figure~\ref{Graph_Energy_Distribution} shows that as the waist of the injected beam increases, the guided energy decreases. The remaining energy is mostly redistributed among the cylindrical modes (modes 1, 6, 15, and 28). This observation aligns with expectations since the impact of the offset becomes less significant for broader beams than for narrower ones. For a large beam waist, the symmetry-breaking effect of a nonzero offset becomes negligible, and only the overlap integrals of cylindrical modes yield significant values.

It is worth noting that the phase also plays a significant role in mode coupling, since the spatial function of the beam in Eq.~\ref{eq:coupling_ceoff} is complex and the mode profiles can have positive and negative regions. In the case of zero transverse offset but a nonzero tilt angle, see the third panel of figure~\ref{Graph_Energy_Distribution}, variations in the tilt angle modify the overlap between the Gaussian input pulse and the propagation modes (see figure~\ref{CI_Modes}.c). This change in overlap affects the energy distribution at the entry of the fiber. As the tilt angle of the injected pulse increases, higher-order modes become more sensitive to the matching of the spatial phase due to their larger transverse spatial extent. Additionally, there is less energy losses compared to the case of an increasing offset since the pulse is still centered. Still, at a high enough input tilt angle, fewer light rays will enter the acceptance cone of the fiber. This cone is directly related to the numerical aperture of the fiber, as shown in figure~\ref{CI_Modes}.e. In this simplified picture, light rays outside this cone fail to undergo total internal reflection and are therefore not guided. Note that the numerical aperture also determines the number of guided modes through the normalized frequency. In this paper, $NA=0.2$ (index contrast 0.0137) meaning that roughly 218 modes are guided.

For a fixed tilt angle, increasing the beam waist shifts the energy distribution toward higher-order modes, as illustrated in the fourth panel of figure~\ref{Graph_Energy_Distribution}. This effect occurs mainly because higher-order modes are larger, and a broader input pulse enhances the corresponding overlap integrals.

\section{Nonlinear modal energy transfer}
\label{sec:BSC}

In the previous section, we demonstrated that the energy distribution at the entry of the fiber can be altered by adjusting the waist and offset of the input pulse. By influencing this initial energy distribution among the propagation modes, it is possible to affect the interactions between these modes during propagation, which occur due to nonlinear processes.

This paper focuses on a specific nonlinear phenomenon unique to multimode fibers, known as beam self-cleaning. Since the context of our work is relatively different from previous work on beam self-cleaning, we prefer to use the term \textit{nonlinear modal energy transfer} to emphasize this difference in framework. Nonlinear modal energy transfer involves a redistribution of energy among the modes during propagation, driven by inter-modal four-wave mixing (IFWM). This process results in a net transfer of energy towards a single dominant mode, causing the transverse profile of the pulse within the fiber to resemble that of the dominant mode. Nonlinear modal energy transfer can thus be interpreted as a restructuring of the transverse profile of the pulse during propagation. While the most common beam self-cleaning outcome is a net transfer of energy towards the fundamental mode, it has also been shown that beam self-cleaning (or nonlinear modal energy transfer) into mode 2 (LP$_{11}$) is possible with simple initial conditions~\cite{deliancourt19-1,chenJiayang22}, and higher modes too with more complex wavefront tuning~\cite{deliancourt19-2}.

In this section, we investigate how the offset and tilt angle of the input pulse influence nonlinear modal energy transfer. The simulations presented here consider a fiber with a core diameter of $50\,\micron$, a numerical aperture $NA=0.2$, and a nonlinear refractive index $n_2=3.2\times 10^{-20}\,V^{-2}m^2$. The fiber length is 30\,cm (with the exception of subsection~\ref{BSCMode4}). The injected pulse is a Gaussian beam of central wavelength 1064\,nm with a full width at half maximum duration of 90\,fs and containing an energy of 38\,nJ, emulating the parameters in Ref.~\cite{liuZ16}. Note that the evolution with distance as a function of input power (i.e. pulse energy and temporal duration), numerical aperture (i.e. fiber diameter and material), and the targeted mode (i.e. initial conditions), changes in a non-trivial way. But we consider a full exploration of that parameter space beyond the scope of this work and not necessarily illuminating at this time.

\subsection{Nonlinear modal energy transfer towards mode 2}
\label{BSCMode2}

\begin{figure}[H]
	\begin{minipage}[c]{.5\linewidth}
		\centering
		\includegraphics[scale=0.5]{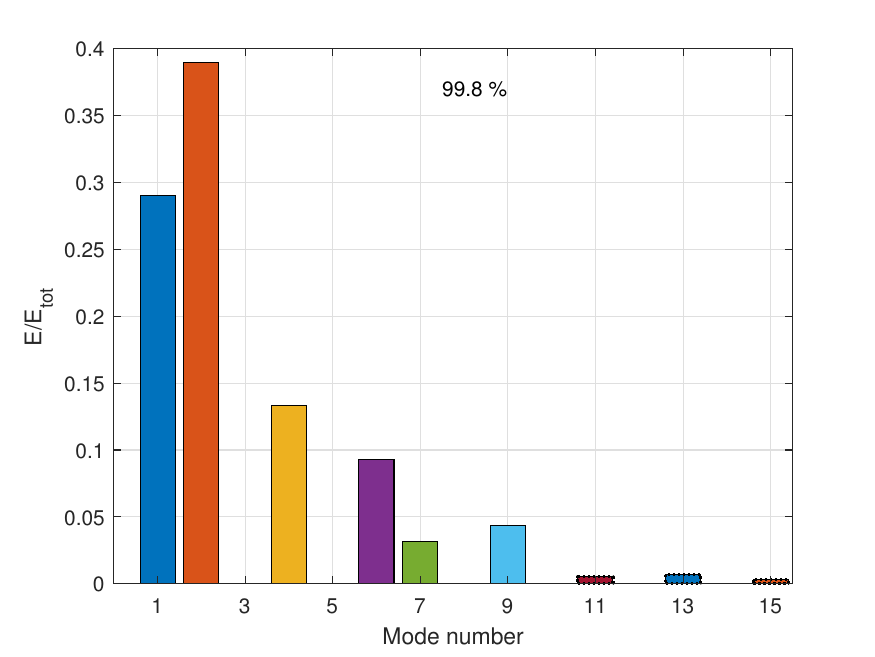}
	\end{minipage}
	\hfill
	\begin{minipage}[c]{.5\linewidth}
		\centering
		\includegraphics[scale=0.5]{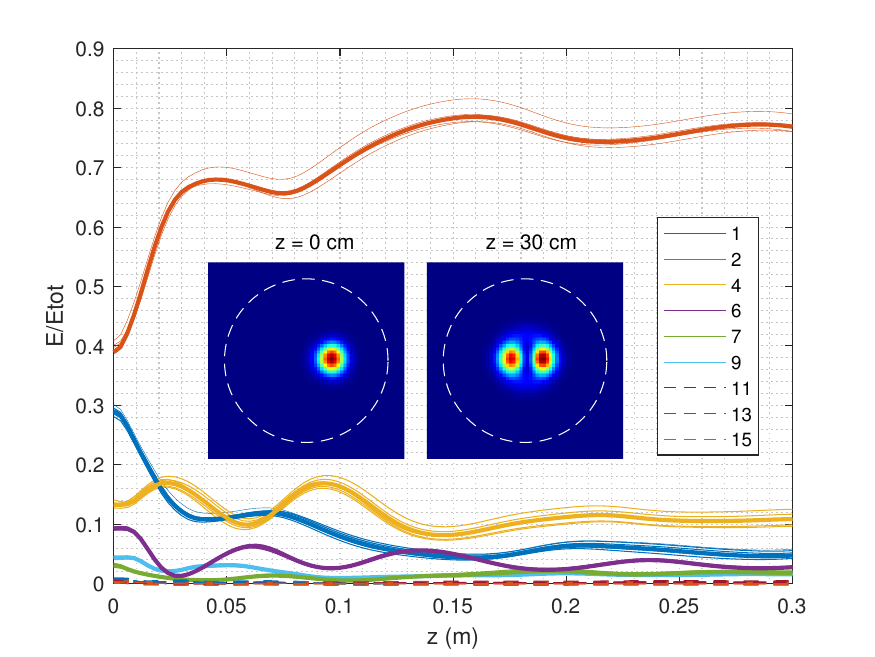}
	\end{minipage}
	\caption{On the left : Initial energy distribution into a basis of 15 propagating modes after injecting a Gaussian beam with a 7\,µm horizontal offset (without noise). The percentage gives the fraction of the energy that is guided by these modes. On the right : Evolution of normalized modal energies versus propagation distance and transverse intensity profile of the pulse-beam at the entry of the fiber (leftmost profile) and after 30\,cm of propagation (rightmost profile). The outline of the fiber core is shown in white dotted lines. Thick lines represent the case of a pulse injected without noise. Thin lines represent the cases where the injected pulse includes noise in both its amplitude and phase. The noise factor is $10\,\%$.}
	\label{Mode2_offset}
\end{figure}

By varying the offset and the waist, we can observe that figure~\ref{Mode2_offset} highlights the specific case where mode 2 (LP$_{11}$) receives the most energy during propagation. This nonlinear modal energy transfer towards mode 2 was achieved with a waist of $6\,\micron$ and an offset of $7\,\micron$ in the $x$-direction. However, the optimal offset value actually does not correspond to the position of the maximum amplitude of mode 2. One might expect that aiming the input beam at the maximum of one of the lobes of mode 2 would maximize energy coupling into this mode. In reality, the offset value that provides the greatest energy gain for mode 2 is higher. This is due to the overlap with the fundamental mode. Indeed, when the input beam is directed towards one of the lobes of mode 2, the Gaussian beam, being larger than the lobe, does not achieve perfect overlap, resulting in a coupling coefficient less than one. Consequently, the remaining energy is distributed among other modes. Since the overlap integral with the fundamental mode is greater than that with mode 2, a significant portion of the energy is coupled into the fundamental mode instead of mode 2.

On the left side of figure~\ref{Mode2_offset}, the energy distribution at the entry of the fiber is depicted. The injected pulse was assumed to be noiseless. Utilizing a small but nonzero offset results in the energy being primarily distributed among lower-order modes. Specifically, approximately $80\,\%$ of the input energy is distributed into modes 1, 2, and 4 (see figure~\ref{CI_Modes}.a). These modes, which account for the majority of the energy coupled into the fiber, are also the most likely to interact with one another, as their propagation constants are similar (see figure~\ref{CI_Modes}.b), resulting in small phase detuning. This enables energy exchange through IFWM over a relatively long distance, leading to a net transfer of energy to mode 2, as shown in the right image of figure~\ref{Mode2_offset}. The outcome is a reshaping, or cleaning, of the transverse profile, revealing the mode 2 profile.

On the right side of figure~\ref{Mode2_offset}, thick lines represent the evolution of modal energies in the case of injecting a noiseless Gaussian pulse-beam. This thus corresponds to the evolution of the initial energy distribution shown on the left side of figure~\ref{Mode2_offset}. Thin lines correspond to cases where Gaussian noise is added to both temporal and spatial amplitude and phase of the input Gaussian pulse-beam. We include five different cases, each varying due to the random nature of the noise, but with a constant noise factor of $10\,\%$. The noise introduces slight variations in the energy distribution among the modes at the entry of the fiber. Nonetheless, figure~\ref{Mode2_offset} shows that the evolution of modal energies is very similar to the case without noise. Hence, our results are robust to noise. In the following, only the noiseless cases will be presented.

Note importantly how the effect we observe differs from traditional beam self-cleaning. Firstly, as already mentioned, we do not model random linear mode coupling. This means that with purely linear propagation (i.e. at low power) the output of the fiber is not highly speckled and the energy content is the same as at the input. However, the output at low power still contains interferences between the modes excited due to their dephasing with propagation. Secondly, because our input conditions are targeting the transfer to specific higher-order modes, we do not see anything similar to the thermalization of the output mode content~\cite{pourbeyram22,baudin23}, where in those cases the \textit{initial} mode content was arbitrary, random, or somehow distributed somewhat equally between a large number of modes (and linear mode coupling contributes). Finally, it has been observed that cleaning into mode 2 may be a transient effect with some parameters~\cite{fabert20}, but we do not observe such an effect, which remains to be investigated more thoroughly.

In a case where a Gaussian pulse with a waist of $10\,\micron$ and a \textit{tilt angle} of 2.5° (and no offset) is injected into the fiber, the energy distribution is relatively similar to the case with an offset, though $98.5\,\%$ of the energy is coupled into the 15 modes compared to $99.8\,\%$ for the case with a nonzero offset. This energy distribution still results in nonlinear modal energy transfer towards mode 2, achieving nearly the same fraction of the total energy after 30\,cm of propagation as in the offset case, though with greater injection losses.

The evolution of the transverse profile during propagation leads to a reshaping of the initial Gaussian profile into the transverse profile of mode 2. While the profile at the entry of the fiber is not perfectly Gaussian, increasing the number of modes improves its reconstruction. This suggests that the 15-mode approximation, while computationally efficient, may not fully capture the complexity of the initial field distribution.

\begin{figure}[H]
	\begin{minipage}[c]{.5\linewidth}
		\centering
		\includegraphics[scale=0.5]{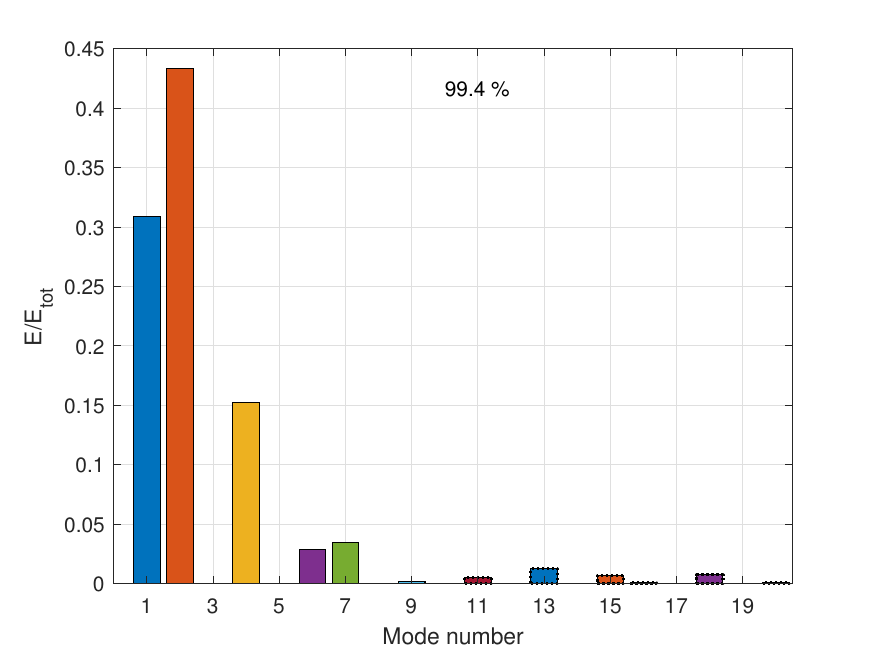}
	\end{minipage}
	\hfill
	\begin{minipage}[c]{.5\linewidth}
		\centering
		\includegraphics[scale=0.5]{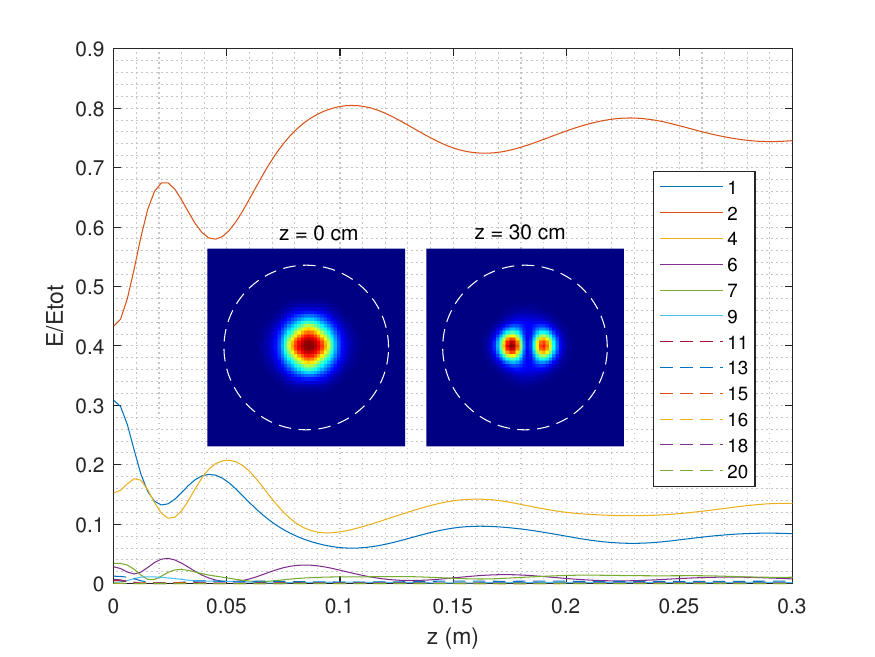}
	\end{minipage}
	\caption{On the left : Initial energy distribution into 20 propagation modes after injecting a tilted Gaussian beam. The percentage gives the fraction of the energy that is contained in those 20 modes. On the right : Evolution of normalized modal energies versus propagation distance and transverse intensity profile of the pulse-beam at the entry of the fiber (leftmost profile) and after 30\,cm of propagation (rightmost profile). The outline of the fiber core is shown in white dotted lines.}
	\label{Mode2_Tilt_20modes}
\end{figure}

When increasing the number of modes to 20, the initial transverse profile more closely resembles a Gaussian, as shown in the leftmost profile of figure~\ref{Mode2_Tilt_20modes}. Additionally, increasing the number of modes enhances energy coupling efficiency at injection since higher-order modes, such as mode 18, now capture a portion of the energy that would otherwise be lost. Despite these refinements, the overall energy redistribution during propagation, primarily governed by IFWM, remains consistent between both cases. After 30\,cm of propagation, the energy in mode 2 differs by only $2.5\,\%$ between the 15- and 20-mode scenarios. This small difference suggests that beyond a certain point, increasing the number of modes yields only marginal improvements in accuracy while significantly increasing computational costs.

For the previous case (see figure~\ref{Mode2_offset}), increasing the number of modes from 15 to 20 has a negligible impact on the results. In the 20-mode case, $100\,\%$ of the energy is coupled into the fiber, compared to $99.8\,\%$ in the 15-mode case. The additional $0.2\,\%$ is distributed among high-order modes, such as modes 16 and 18, and has a negligible impact on the nonlinear phenomena occurring during propagation. Even when considering only 10 modes, the results remain satisfactory, as the energy in mode 2 after 30\,cm of propagation differs by only $7\,\%$. This demonstrates that even a reduced number of modes can provide meaningful insights while significantly lowering computational costs.

In conclusion, when a pulse is injected with a nonzero tilt angle, more modes are excited compared to the no-tilt case. Consequently, the number of modes considered in the simulation should be adjusted to account for these additional contributions.

\subsubsection{Discussion about nonlinear modal energy transfer towards into mode 2}
Figures~\ref{Mode2_offset} and~\ref{Mode2_Tilt_20modes} show that the transverse profile of the pulse-beam can be reshaped into the transverse profile of mode 2. It is also worth noting that the remaining modes retain a small fraction of their energy, contributing to an increase of the entropy of the overall system. For instance, in the case of figure~\ref{Mode2_offset}, the remaining modes (modes 1, 4, 6, 7, and 9) collectively retain approximately $22\,\%$ of the initial energy. These modes form a noisy background upon which the transverse profile of mode 2 emerges. This background noise is not a problem, because what is of interest here for applications is beam intensity. Since the intensity of the lobes is much greater than that of the noise, nonlinear effects will only occur in the spatial region occupied by the lobes with the greatest intensity.

These high-intensity lobes can be thought of as nonlinear waveguides, since they locally increase the value of the refractive index. The quality of these waveguides depends directly on the evolution of modal energies. In the case of figure~\ref{Mode2_offset}, by inspecting the transverse profile at different points of propagation, we see that energy is well distributed in both lobes leading to two optical waveguides. On the other hand, in the case of figure~\ref{Mode2_Tilt_20modes}, the energy tends to be more distributed into one lobe so only one optical waveguide can be considered. This is still interesting since a lobe of mode 2 is smaller than the lobe of the fundamental mode.

\subsection{Nonlinear modal energy transfer towards mode 4}
\label{BSCMode4}

\begin{figure}[H]
	\begin{minipage}[c]{.5\linewidth}
		\centering
		\includegraphics[scale=0.5]{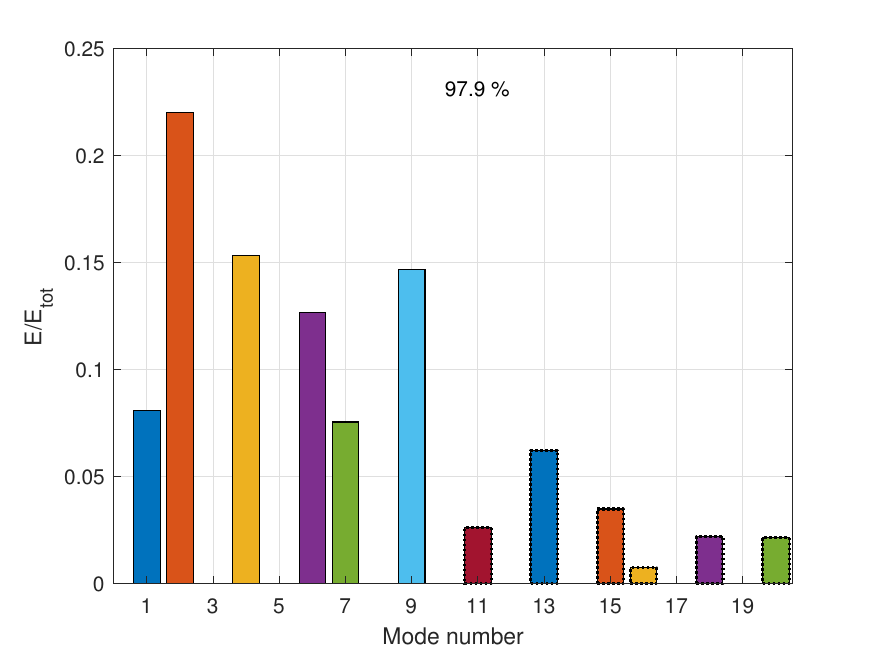}
	\end{minipage}
	\hfill
	\begin{minipage}[c]{.5\linewidth}
		\centering
		\includegraphics[scale=0.5]{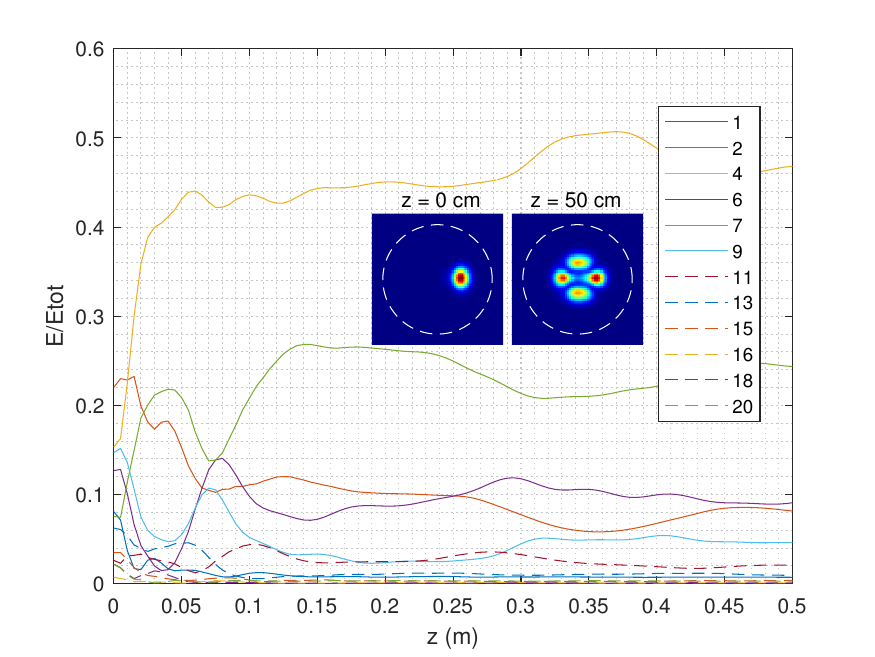}
	\end{minipage}
	\caption{On the left : Initial energy distribution into 20 propagation modes after injecting a Gaussian beam with a 10 µm horizontal offset. The percentage gives the fraction of the energy that is guided in those 20 modes. On the right : Evolution of normalized modal energies versus propagation distance and transverse intensity profile of the pulse-beam at the entry of the fiber (leftmost profile) and after 50 cm of propagation (rightmost profile). The outline of the fiber core is shown in white dotted lines.}
	\label{Mode4_offset_20modes}
\end{figure}

In this subsection, we demonstrate that by using a larger offset, higher-order modes can be excited, leading to nonlinear modal energy transfer towards a mode previously inaccessible with such simple initial conditions. By using a waist of $6\,\micron$ and an offset of $10\,\micron$, nonlinear modal energy transfer towards mode 4 (LP$_{21}$) is achieved, as shown in the figure~\ref{Mode4_offset_20modes}. The left panel of this figure illustrates the energy distribution at the fiber entry. In this subsection, a fiber length of 50\,cm is considered. A larger offset than in the previous subsection is used because the lobes of mode 4 are located farther from the fiber center than those of mode 2. As before, targeting a lobe of mode 4 with the input pulse requires an offset greater than the position of the maximum of the lobe. This is due to the influence of the fundamental mode as well as mode 2.

For the simulation, 20 modes were considered because injecting the Gaussian pulse with a higher offset excites a larger number of modes. It is therefore necessary to include enough modes to ensure that the energy distribution is accurately represented. A simulation with 25 modes was also performed, but no significant differences were observed compared to the 20-mode case, indicating that the latter is a reasonable approximation. In the 25-mode simulation, $98.6\,\%$ of the energy was injected into the multimode fiber, compared to $97.9\,\%$ for the 20-mode case. The $0.7\,\%$ difference was distributed between modes 22 and 24, which carry too little energy to significantly impact the results. Consequently, these higher-order modes can be neglected without compromising the accuracy of the simulation. In this case, the fiber length is increased up to $50$ cm. This increasing of the fiber length is motivated by energy exchange between mode 4 and mode 7 that are still important after 30\,cm of propagation.

The left panel shows that, for this offset, the energy is primarily distributed among modes 2 through 9 (see figure~\ref{CI_Modes}.a for mode numbering). During propagation, modes 4 and 7 gain energy via inter-modal four-wave mixing. Mode 4 quickly becomes the dominant mode, although mode 2 initially carried the most energy. Consequently, the transverse profile of the pulse-beam is reshaped to match that of mode 4. However, mode 7 retains approximately $24\,\%$ of the initial energy, which leads to fluctuations in the transverse profile during propagation due to interference between modes 4 and 7.

\subsubsection{Discussion about nonlinear modal energy transfer towards mode 4}
Figure~\ref{Mode4_offset_20modes} shows the transverse profile of the beam at the output of the fiber. During propagation, the transverse profile was restructured from a single Gaussian lobe into four distinct lobes. However, when compared with the transverse profile of mode 4 shown in figure~\ref{CI_Modes}, it becomes evident that, while the four lobes are separated, they are also distorted due to interference with mode 7. To achieve improved nonlinear modal energy transfer towards mode 4, mode 7 would need to transfer its energy to mode 4. In such a case, the transverse profile would experience significantly fewer fluctuations during propagation through the multimode fiber and lobes would be more symmetric. In the case illustrated in figure~\ref{Mode4_offset_20modes}, two of the lobes exhibit sufficiently high intensity to function as nonlinear optical waveguides, for example.

\subsection{Nonlinear modal energy transfer towards mode 7}
\label{BSCM7}

\begin{figure}[H]
	\begin{minipage}[c]{.5\linewidth}
		\centering
		\includegraphics[scale=0.5]{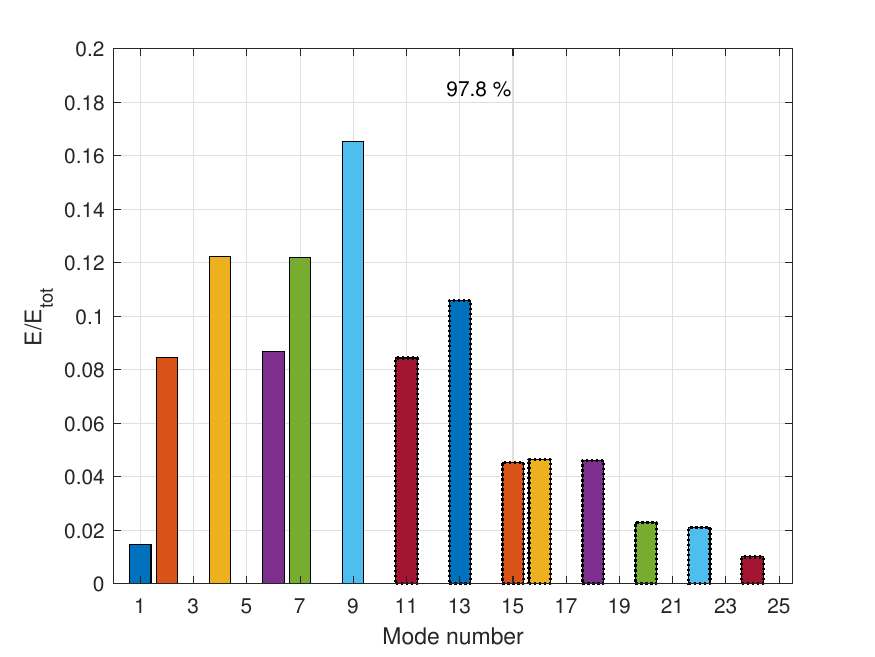}
	\end{minipage}
	\hfill
	\begin{minipage}[c]{.5\linewidth}
		\centering
		\includegraphics[scale=0.5]{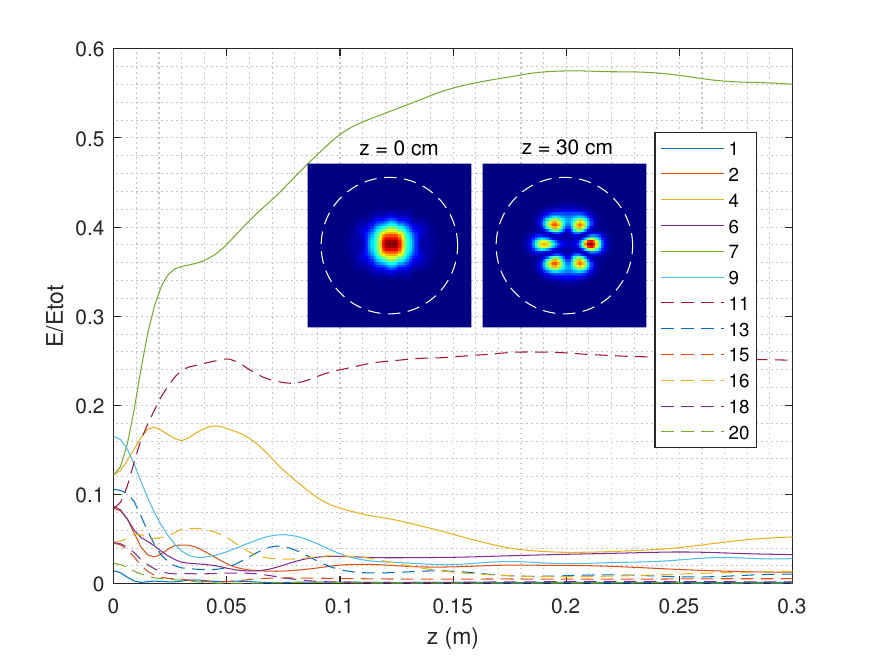}
	\end{minipage}
	\caption{On the left : Initial energy distribution into 25 propagation modes after injecting a Gaussian beam with a $5^{\circ}$ tilt angle. The percentage gives the fraction of the energy that is guided in those 25 modes. On the right : Evolution of normalized modal energies versus propagation distance and transverse intensity profile of the pulse-beam at the entry of the fiber (leftmost profile) and after 30\,cm of propagation (rightmost profile). The outline of the fiber core is shown in white dotted lines.}
	\label{Mode7_Tilt_25modes}
\end{figure}

As in the previous subsection, we demonstrate here that varying one of the injection parameters allows for the observation of a nonlinear modal energy transfer towards another new higher-order mode, not previously accessible with these simple initial conditions. As shown in figure~\ref{Mode7_Tilt_25modes}, injecting a Gaussian tilted pulse-beam with a waist of $10\,\micron$ and a tilt angle of 5° (see figure~\ref{CI_Modes}.c) results in a nonlinear modal energy transfer towards mode 7 (LP$_{31}$). In this case, most of the energy at the fiber entry is distributed among modes 4 through 13, with mode 9 being the dominant mode initially. By comparing the left panels of figures~\ref{Mode2_Tilt_20modes} and~\ref{Mode7_Tilt_25modes}, it is evident that increasing the tilt angle leads to the excitation of a significantly larger number of modes (for a fixed waist), as mentioned in section~\ref{sec:mode_coupling}. This implies that the number of modes considered in simulations must increase with the tilt angle. We can see that the transverse profile of the beam at the very beginning of the fiber does not look like a perfect bell shape. Increasing the number of modes to be considered in the basis to 30 improves the transverse profile, but new modes contain less than $1\%$ of the total energy so they are negligible in the propagation simulation. Furthermore, increasing the waist results in more energy being coupled into higher-order modes, as shown in the fourth panel of figure~\ref{Graph_Energy_Distribution}.

Note that is also possible to obtain nonlinear modal energy transfer towards mode 7 by using no tilt, but a small waist (e.g. $6\,\micron$) with a large transverse offset (e.g. $13\,\micron$). Nevertheless, in such a case, mode 7 gains generally less energy than in a case with nonzero tilt. It is also important to note that the method used to aim at a lobe of the mode of interest is not efficient for mode 7. This inefficiency arises because the lobes of mode 7 are relatively small and located far from the center of the fiber. Additionally, the influence of lower-order modes must be taken into account when targeting one of the lobes of mode 7. This requires considering a higher offset than the position of the lobes themselves, as explained in subsection~\ref{BSCMode2}. However, this approach necessitates using a very small waist and a large offset, which are not realistic experimentally. A large offset leads to significant losses, while an extremely small waist is impractical to achieve in an experimental setup.

As shown in figure~\ref{Mode7_Tilt_25modes}, a net transfer of energy occurs during propagation, primarily toward modes 7 and 11. Mode 7 rapidly becomes the dominant mode, while the energy in mode 11 saturates earlier than in mode 7. Consequently, the transverse profile of the Gaussian beam is reshaped to match that of mode 7. Mode 11, carrying approximately $25\,\%$ of the incident pulse energy, has enough energy compared to mode 7 to create significant interference with it. This interference results in a transverse profile that may fluctuate during propagation.

\subsubsection{Discussion about nonlinear modal energy transfer towards mode 7}
As described before, the transverse profile of mode 7 emerges during propagation. Specifically, the Gaussian lobe at the fiber entry is subdivided into six (nearly) distinct lobes that characterize the transverse profile of mode 7. However, when compared to the transverse profile of mode 7 shown in figure~\ref{CI_Modes}.a, we see that the lobes are slightly deformed, and some are not entirely separated from their neighbors. This deformation arises from interference caused by higher-order modes—--in this case, mode 11—--which carries enough energy to alter the lobes.

In an ideal scenario, mode 11 would transfer part of its energy to mode 7, eventually reaching the same low energy level as the other higher-order modes. In such a case, the transverse profile would remain more stable throughout propagation. The influence of higher-order modes would be minimized, resulting in well-separated and symmetrical lobes that are free from deformation. Lobes that are stable through propagation could be used as nonlinear optical waveguides as mentioned for the previous cases.

\section{Spatial chirp}
\label{sec:SC}

In this section, we consider an input light pulse-beam that is space-time coupled, meaning that the spatial and temporal variables in the equation characterizing the pulse are not separable~\cite{akturk10}. Physically, this coupling leads to a distinct spatial behavior for each frequency component of the pulse. In other words, when the pulse is injected into the multimode fiber, each frequency component enters with different initial conditions. This phenomenon is not accounted for by the coupling coefficient presented in Eq.~\ref{eq:coupling_ceoff}, necessitating the use of the following coupling coefficient~\cite{jolly23-2}:

\begin{equation}
	\mu_i(\omega) = \frac{\int\int \tilde{A}(x,y,\Delta z,\omega)F_i(x,y)dxdy}{\sqrt{\int\int\left|\tilde{A}(x,y,\Delta z,\omega)\right|^2dxdyd\omega \int\int\left|F_i(x,y)\right|^2dxdy}}.
\end{equation}

\noindent This coefficient, which measures the coupling of the input field into each mode, is a frequency-dependent complex number. Consequently, $\mu_i$ represents a complex frequency envelope rather than an efficiency, as $\eta_i$ does. The coupling efficiency into mode \textit{i} is given by $\int\left|\mu_i(\omega)\right|^2d\omega$. Maybe more intuitively, a spatial-temporal coupling results in each guided mode having a different temporal profile, a direct result of the Fourier-transform of the frequency-dependent coupling coefficient $\mu_i(\omega)$.

In this paper, the space-time coupling studied is the so-called spatial chirp~\cite{gu04}. In this case, each frequency component of the pulse has a different transverse offset at the focus point of the beam. In the time domain, spatial chirp corresponds to a time-dependent wavefront tilt (called wavefront rotation), which has been used for attosecond physics experiments~\cite{quere14}. To create spatial chirp, one can focus a beam with angular dispersion (a beam with tilted pulse front in the time domain) that can be created by passage through prisms or gratings. To model spatial chirp, the frequency distribution along the transverse direction is described by the frequency-dependent transverse offset $x_0(\omega)=w_0\tau_t\delta\omega / 2$, where $w_0$ is the beam waist, $\delta\omega=\omega-\omega_0$ is the frequency offset relative to the central frequency of the pulse, and $\tau_t$ is the amount of pulse-front tilt applied to the collimated beam to generate spatial chirp in the focused beam. The parameter $\tau_t$ essentially determines the strength of the spatial chirp.

\begin{figure}[H]
	\begin{minipage}[c]{.55\linewidth}
		\centering
		\begin{overpic}[scale=0.3]{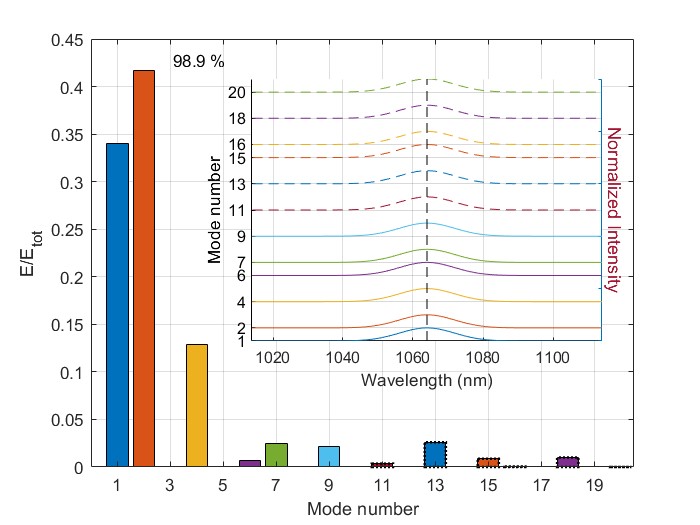}
			\put(-5,140){a)}
		\end{overpic}
	\end{minipage}
	\hfill
	\begin{minipage}[c]{.55\linewidth}
		\centering
		\begin{overpic}[scale=0.3]{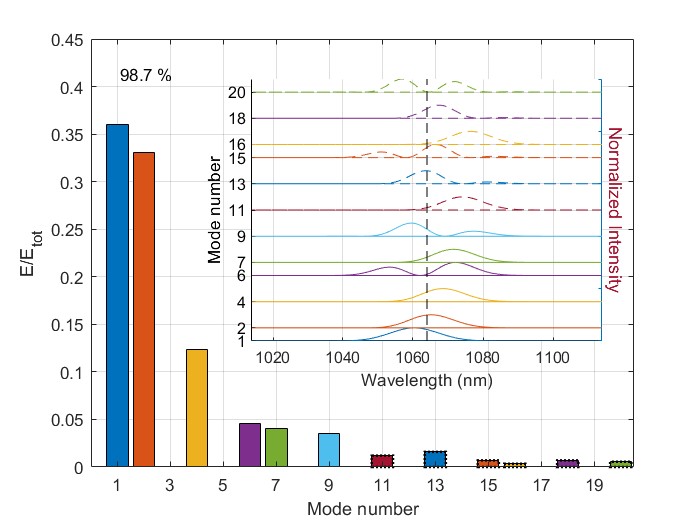}
			\put(0,140){b)}
		\end{overpic}
	\end{minipage}
	\hfill
	\begin{minipage}[c]{.55\linewidth}
		\centering
		\begin{overpic}[scale=0.5]{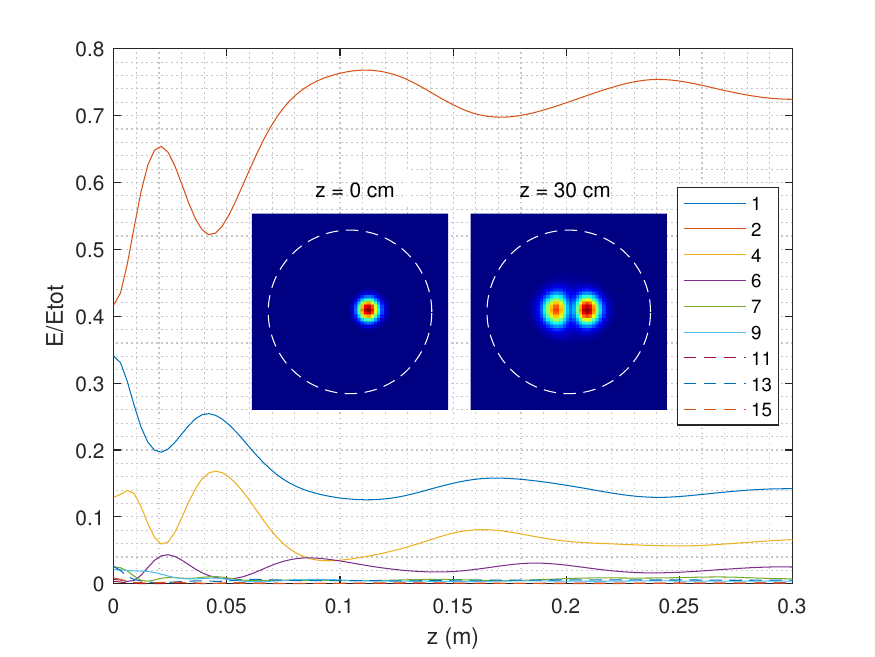}
			\put(-5,140){c)}
		\end{overpic}
	\end{minipage}
	\hfill
	\begin{minipage}[c]{.55\linewidth}
		\centering
		\begin{overpic}[scale=0.5]{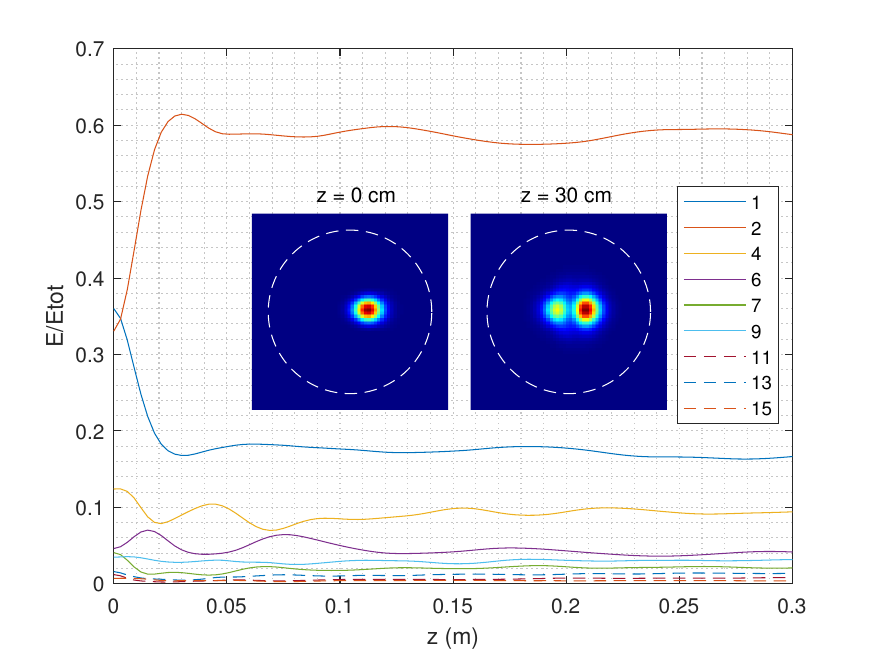}
			\put(0,140){d)}
		\end{overpic}
	\end{minipage}
	\caption{a) Energy distribution among 15 modes at the entry of the multimode fiber and spectra of the excited modes in a case without spatial chirp. The black vertical line indicates the central wavelength of the input pulse. b) Energy distribution among 15 modes at the entry of the multimode fiber and spectra of the excited modes in a case with spatial chirp. The black vertical line indicates the central wavelength of the input pulse. c) d) In a case without and with spatial chirp respectively : Evolution of normalized modal energies versus propagation distance and transverse intensity profile of the pulse-beam. The outline of the fiber core is shown in white dotted lines.}
	\label{Comparaison_spatial_chirp}
\end{figure}

In the following analysis, we consider the injection of a Gaussian pulse-beam with a waist of 4\,µm and a transverse offset in the x-direction of 5\,µm. Fiber parameters as well as pulse energy and its duration are the same as in section~\ref{sec:BSC}. In one scenario, there is no spatial chirp ($\tau_t=0$), while in the other, $\tau_t=\tau_0$, where $\tau_0 = \text{FWHM}/\sqrt{2\ln(2)}$. Figure~\ref{Comparaison_spatial_chirp} shows energy distribution among modes at the beginning of the multimode fiber in a case without spatial chirp (figure~\ref{Comparaison_spatial_chirp}.a) and in the case with spatial chirp (figure~\ref{Comparaison_spatial_chirp}.b). It also shows the evolution of modal energies for the case without spatial chirp (figure~\ref{Comparaison_spatial_chirp}.c) and the case with it (figure~\ref{Comparaison_spatial_chirp}.d). In both cases, we have a nonlinear modal energy transfer towards mode 2. 

First, spatial chirp impacts the energy distribution among modes, but it also changes the spectrum of each mode individually, as shown in the figure~\ref{Comparaison_spatial_chirp}.a and~\ref{Comparaison_spatial_chirp}.b, leading to a new degree of freedom to impact nonlinear interactions between modes during propagation. In the case with spatial chirp, the fundamental mode initially carries more energy than mode 2. Additionally, $98.7\,\%$ of the total energy is coupled into the 15 modes of the multimode fiber that we consider, which is slightly less than in the case without spatial chirp. By the end of propagation, mode 2 retains less energy than in the case without spatial chirp, but it has also begun propagation with a lower initial energy.

\begin{figure}[H]
	\begin{minipage}[c]{.5\linewidth}
		\centering
		\begin{overpic}[scale=0.4]{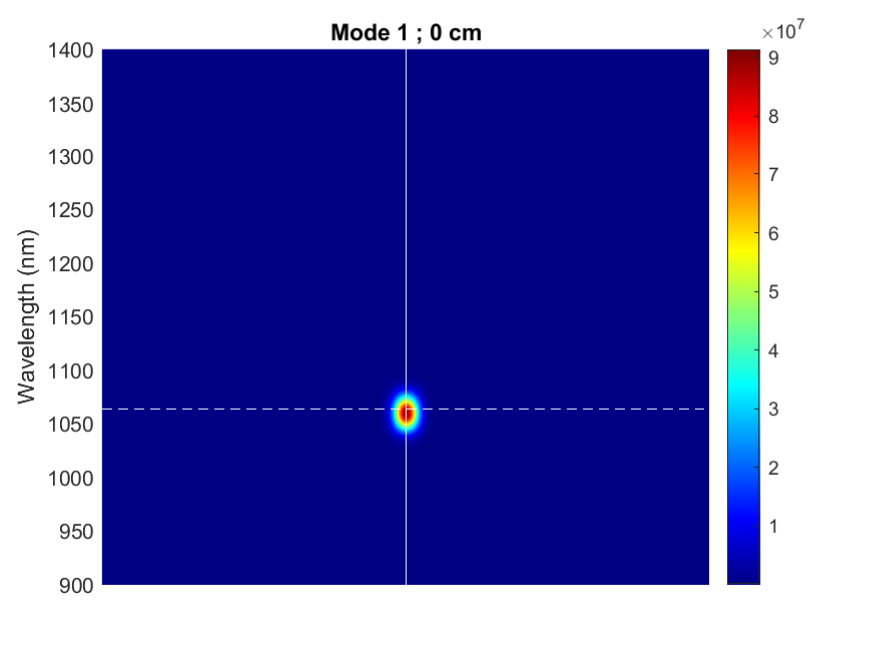}
			\put(-5,115){a)}
		\end{overpic}
	\end{minipage}
	\hspace{-3em}
	\begin{minipage}[c]{.475\linewidth}
		\centering
		\begin{overpic}[scale=0.4]{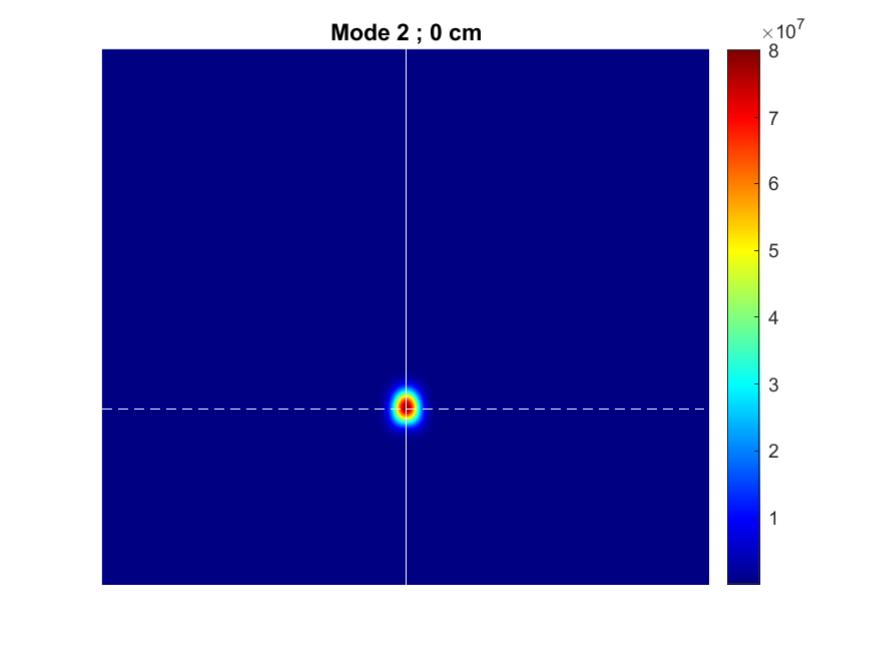}
			\put(10,115){d)}
		\end{overpic}
	\end{minipage}
	\hfill
	% Espace réduit entre les lignes
	\vspace{-0.4cm}
	
	\begin{minipage}[c]{.5\linewidth}
		\centering
		\begin{overpic}[scale=0.4]{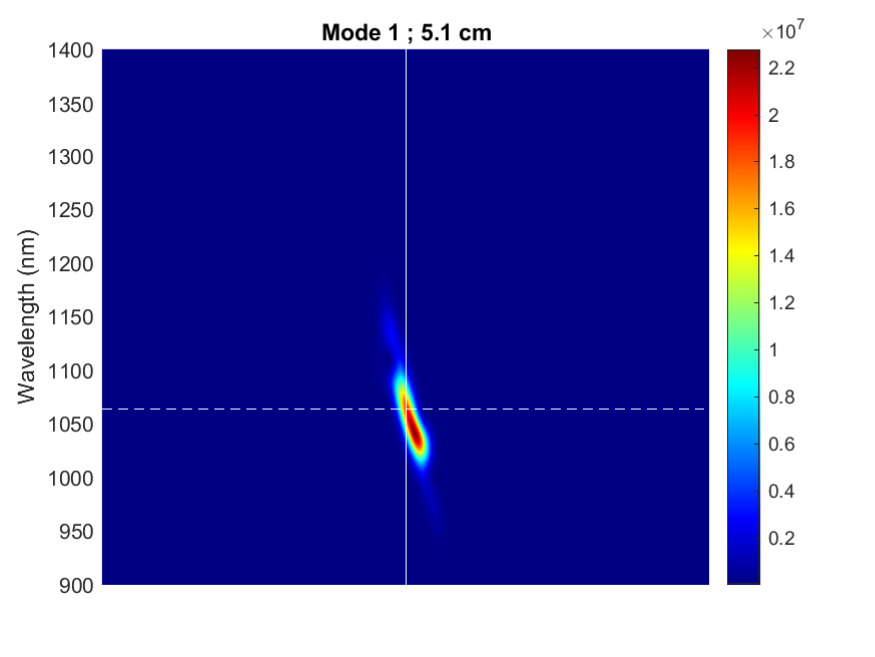}
			\put(-5,115){b)}
		\end{overpic}
	\end{minipage}
	\hspace{-3.5em}
	\begin{minipage}[c]{.5\linewidth}
		\centering
		\begin{overpic}[scale=0.4]{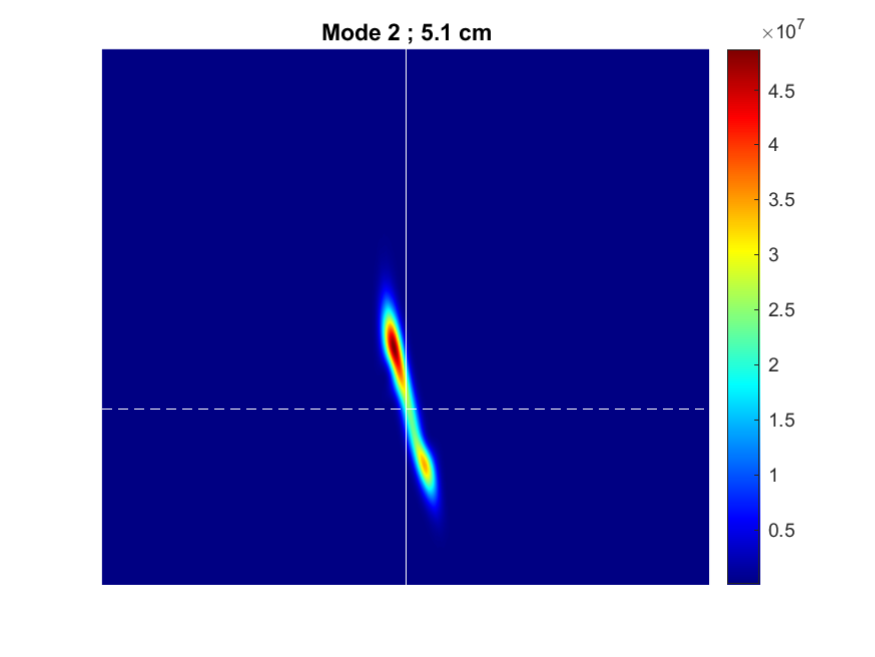}
			\put(10,115){e)}
		\end{overpic}
	\end{minipage}
	\hfill
	% Espace réduit entre les lignes
	\vspace{-0.4cm}
	
	\begin{minipage}[c]{.5\linewidth}
		\centering
		\begin{overpic}[scale=0.4]{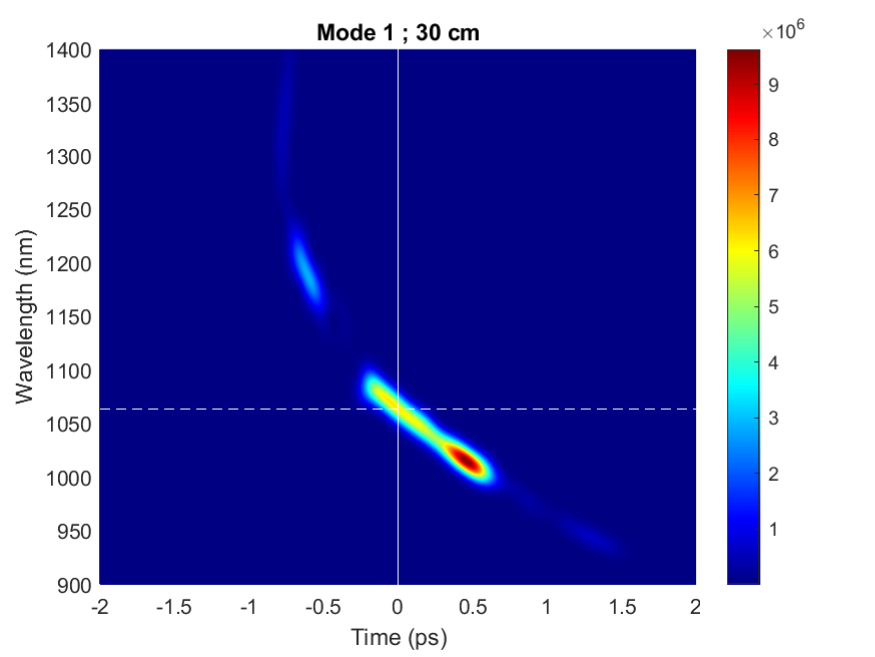}
			\put(-5,115){c)}
		\end{overpic}
	\end{minipage}
	\hspace{-3.5em}
	\begin{minipage}[c]{.51\linewidth}
		\centering
		\begin{overpic}[scale=0.4]{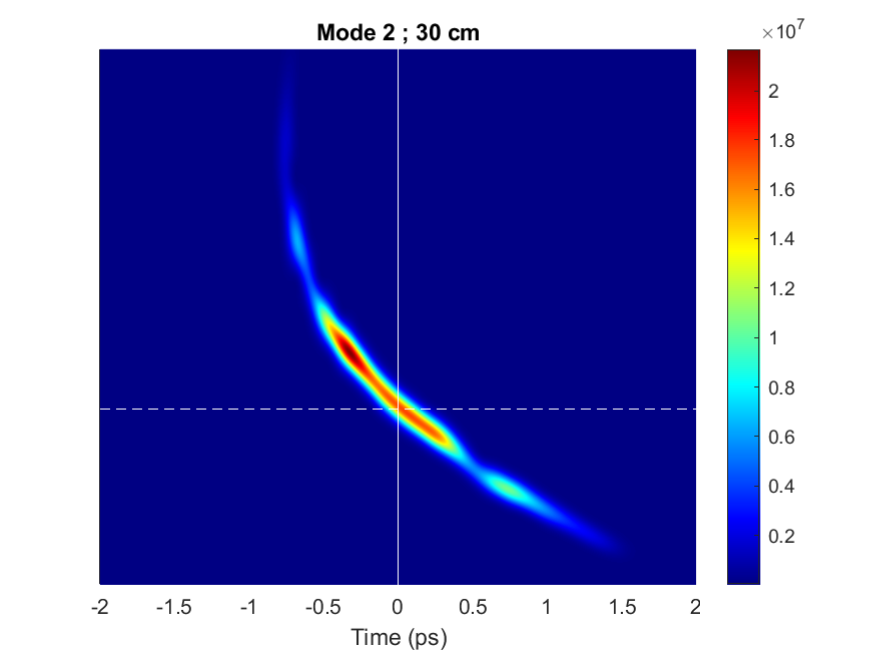}
			\put(10,115){f)}
		\end{overpic}
	\end{minipage}
	\caption{a) b) c) Spectrograms of mode 1 after 0 cm, 5.1 cm and 30 cm of propagation, respectively. d) e) f) Spectrograms of mode 2 after 0\,cm, 5.1\,cm and 30\,cm of propagation, respectively. The white dotted line indicates the central wavelength of the input pulse. The white vertical line indicates time zero.}
	\label{Spectrogrammes}
\end{figure}

Furthermore, in the presence of spatial chirp, the energy of mode 2 fluctuates significantly less during propagation. This change in modal energy evolution arises because the modes have different spectra at the input of the fiber (see figure~\ref{Spectrogrammes}.a and~\ref{Spectrogrammes}.d), altering the nonlinear interactions between modes that govern their energy exchange.

In the case of a positive spatial chirp combined with a positive transverse offset, figure~\ref{Comparaison_spatial_chirp}.b shows that the higher the mode order, the more its spectrum is centered at longer wavelengths. Figure~\ref{Spectrogrammes} presents spectrograms of modes 1 and 2 at different points along the fiber. At the beginning of propagation, the spectra of the modes are slightly offset from the central frequency of the incident pulse (1064\,nm). The spectrum of mode 2 is slightly shifted toward longer wavelengths (see figure~\ref{Spectrogrammes}.b), while the spectrum of the fundamental mode is shifted toward shorter wavelengths (see figure~\ref{Spectrogrammes}.a). Higher-order modes generally exhibit spectra shifted toward longer wavelengths. This spectral shift also affects the temporal shapes of the modes, making them temporally broader compared to the case without spatial chirp. After 5.1\,cm of propagation (see figures~\ref{Spectrogrammes}.b and~\ref{Spectrogrammes}.e), the spectrum of the fundamental mode broadens and shifts to shorter wavelengths, with most of its spectrum lying between 1016\,nm and 1090\,nm. The spectrum of mode 2, on the other hand, splits into two distinct lobes: a primary lobe at approximately 1121\,nm and a secondary lobe at around 1010\,nm. While some frequencies exist between these lobes, they constitute only a minor part of the spectrum.

At the end of the fiber (see figures~\ref{Spectrogrammes}.c and~\ref{Spectrogrammes}.f), the spectrum of the fundamental mode consists of a primary lobe centered around 1016\,nm, followed by a tail extending from 1032\,nm to 1088\,nm. In contrast, the spectrum of mode 2 shifts further towards longer wavelengths, spanning primarily between 1020\,nm and 1062\,nm, with a small lobe centered around 989\,nm.

\subsection{Discussion}

The difference in modal energy evolution between cases with and without spatial chirp arises from the distinct evolution of the mode spectra. In the case without spatial chirp, the fundamental mode and mode 2 exhibit a similar spectral evolution during the first centimeters of propagation. Because they share comparable frequency components, they propagate at nearly the same speed and they have a small phase mismatch that facilitates significant energy exchange through inter-modal four-wave mixing.

However, this behavior changes in the presence of spatial chirp. Figures~\ref{Spectrogrammes}.b and~\ref{Spectrogrammes}.e show that after 5.1\,cm of propagation, the spectra of modes 1 and 2 have different shapes, sharing fewer overlapping points on the time grid. At 5.1\,cm, the bulk of the fundamental mode spectrum no longer coincides temporally with the spectrum of mode 2. The latter is now primarily composed of two lobes: one at a lower frequency than the fundamental mode spectrum and the other at a higher frequency. Due to chromatic dispersion, these lobes propagate at different speeds, leading to a temporal separation (imperfect overlap) between the fundamental mode and mode 2. This separation prevents the two modes from exchanging energy via IFWM after this point, since it is an instantaneous nonlinear process. Moreover, as the spectra of the modes increasingly differ in frequency content, the phase mismatch grows, further reducing the potential for energy exchange. The small fluctuations observed during propagation are due to the modes maintaining limited spectral overlap. As shown in figure~\ref{Spectrogrammes}.c, the tail of the fundamental mode spectrum shares certain wavelengths with mode 2, allowing for minor energy exchange.

In sum, the above phenomena cause the evolution of the exchange of energy between mode 1 and mode 2 to essentially finish after a shorter propagation distance in the fiber when spatial chirp is present. The oscillations in the energy of mode 2, which characterize an exchange of energy (beating) between mode 1 and mode 2, are no longer as strong. Such a control is enabled by the spatial chirp and is not necessarily possible by controlling another parameter such as initial energy. 

\section{Conclusion}
\label{sec:conclusion}

In conclusion, our study presents a comprehensive numerical analysis of the influence of experimentally accessible initial conditions on nonlinear interactions during the propagation of a femtosecond optical pulse in a multi-mode fiber. By adjusting the transverse offset, waist, and tilt angle of the input Gaussian pulse-beam, we demonstrated that the energy distribution among the modes at the entry of the fiber can be effectively controlled using experimentally accessible injection parameters.

Furthermore, our numerical simulations revealed that these simple initial conditions enable the on-demand reshaping of a Gaussian beam into the spatial profiles of higher-order modes, specifically modes 2, 4, and 7 (LP$_{11}$, LP$_{21}$, and LP$_{31}$, respectively). We could envision, for example, a system for converting from one mode to another by considering two fibers that are aligned but with slightly offset centers. The Gaussian pulse-beam considered would be the fundamental mode emerging from the first fiber. This system would serve as an alternative to those based on the use of spatial light modulators, which are complex to use and result in significant losses.

Additionally, we introduced a new degree of freedom by injecting a space-time coupled pulse. We focused on a relatively simple form of space-time coupling, known as spatial chirp, which allows different spatial modes to exhibit distinct temporal and spectral profiles. This significantly affects nonlinear modal energy transfer processes based on inter-modal four-wave mixing. Especially important in the case of spatial chirp and space-time coupled beams in general, is that they provide a path towards new dynamics not accessible simply by changing pulse energy (or duration).

Our findings pave the way for advanced control of multi-mode nonlinear optical dynamics by exploiting a high-dimensional space of space-time initial conditions, especially those that are experimentally simple, offering promising avenues for future research and applications in the field.

\begin{backmatter}
	
\bmsection{Funding}
Fonds De La Recherche Scientifique - FNRS.

\bmsection{Disclosures}
The authors declare no conflicts of interest.

\bmsection{Data Availability Statement}
Data underlying the results presented in this paper are not publicly available at this time but may be obtained from the authors upon reasonable request.

\end{backmatter}

%%%%%%%%%%%%%%%%%%%%%%% References %%%%%%%%%%%%%%%%%%%%%%%%%


\begin{thebibliography}{10}
	\newcommand{\enquote}[1]{``#1''}
	
	\bibitem{cizmar12}
	T.~Čižmár and K.~Dolakia, \enquote{Exploiting multimode waveguides for pure
		fibre-based imaging,} {\protect\JournalTitle{Nature Communications}}
	\textbf{3}, 1027 (2012).
	
	\bibitem{ploeschner15}
	M.~Plöschner, T.~Tyc, and T.~Čižmár, \enquote{Seeing through chaos in
		multimode fibres,} {\protect\JournalTitle{Nature Photonics}} \textbf{9},
	529--535 (2015).
	
	\bibitem{caoH23}
	H.~Cao, T.~\v{C}i\v{z}m\'{a}r, S.~Turtaev, \emph{et~al.}, \enquote{Controlling
		light propagation in multimode fibers for imaging, spectroscopy, and beyond,}
	{\protect\JournalTitle{Advances in Optics and Photonics}} \textbf{15},
	524--612 (2023).
	
	\bibitem{krupa19}
	K.~Krupa, A.~Tonello, A.~Barthélémy, \emph{et~al.}, \enquote{Multimode
		nonlinear fiber optics, a spatiotemporal avenue,} {\protect\JournalTitle{APL
			Photonics}} \textbf{4}, 110901 (2019).
	
	\bibitem{wright22-1}
	L.~G. Wright, W.~H. Renninger, D.~N. Christodoulides, and F.~W. Wise,
	\enquote{Nonlinear multimode photonics: nonlinear optics with many degrees of
		freedom,} {\protect\JournalTitle{Optica}} \textbf{9}, 824--–841 (2022).
	
	\bibitem{wright22-2}
	L.~G. Wright, F.~O. Wu, D.~N. Christodoulides, and F.~W. Wise, \enquote{Physics
		of highly multimode nonlinear optical systems,} {\protect\JournalTitle{Nature
			Physics}} \textbf{18}, 1018--1030 (2022).
	
	\bibitem{liuZ16}
	Z.~Liu, L.~G. Wright, D.~N. Christodoulides, and F.~W. Wise, \enquote{Kerr
		self-cleaning of femtosecond-pulsed beams in graded-index multimode fiber,}
	{\protect\JournalTitle{Optics Letters}} \textbf{41}, 3675--3678 (2016).
	
	\bibitem{krupa17}
	K.~Krupa, A.~Tonello, B.~M. Shalaby, \emph{et~al.}, \enquote{Spatial beam
		self-cleaning in multimode fibres,} {\protect\JournalTitle{Nature Photonics}}
	\textbf{11}, 237--242 (2017).
	
	\bibitem{pourbeyram22}
	H.~Pourbeyram, P.~Sidorenko, F.~O. Wu, \emph{et~al.}, \enquote{Direct
		observations of thermalization to a rayleigh--jeans distribution in multimode
		optical fibres,} {\protect\JournalTitle{Nature Physics}} \textbf{18},
	685--690 (2022).
	
	\bibitem{baudin23}
	K.~Baudin, J.~Garnier, A.~Fusaro, \emph{et~al.}, \enquote{Observation of light
		thermalization to negative-temperature rayleigh-jeans equilibrium states in
		multimode optical fibers,} {\protect\JournalTitle{Physical Review Letters}}
	\textbf{130}, 063801 (2023).
	
	\bibitem{deliancourt19-1}
	E.~Deliancourt, M.~Fabert, A.~Tonello, \emph{et~al.}, \enquote{Kerr beam
		self-cleaning on the lp11 mode in graded-index multimode fibers,}
	{\protect\JournalTitle{OSA Continuum}} \textbf{2}, 1089--1096 (2019).
	
	\bibitem{deliancourt19-2}
	E.~Deliancourt, M.~Fabert, A.~Tonello, \emph{et~al.}, \enquote{Wavefront
		shaping for optimized many-mode kerr beam self-cleaning in graded-index
		multimode fiber,} {\protect\JournalTitle{Optics Express}} \textbf{27},
	17311--17321 (2019).
	
	\bibitem{chenJiayang22}
	J.~Chen, W.~Hong, and A.~Luo, \enquote{Nonlinear dynamics of beam self-cleaning
		on {LP}{11} mode in multimode fibers,} {\protect\JournalTitle{Optics
			Express}} \textbf{30}, 43453--43463 (2022).
	
	\bibitem{wright15-2}
	L.~G. Wright, D.~N. Christodoulides, and F.~W. Wise, \enquote{Controllable
		spatiotemporal nonlinear effects in multimode fibres,}
	{\protect\JournalTitle{Nature Photonics}} \textbf{9}, 306--310 (2015).
	
	\bibitem{wright15-3}
	L.~G. Wright, S.~Wabnitz, D.~N. Christodoulides, and F.~W. Wise,
	\enquote{Ultrabroadband dispersive radiation by spatiotemporal oscillation of
		multimode waves,} {\protect\JournalTitle{Physical Review Letters}}
	\textbf{115}, 223902 (2015).
	
	\bibitem{eftekhar17}
	M.~A. Eftekhar, L.~G. Wright, M.~S. Mills, \emph{et~al.}, \enquote{Versatile
		supercontinuum generation in parabolic multimode optical fibers,}
	{\protect\JournalTitle{Optics Express}} \textbf{25}, 9078--9087 (2017).
	
	\bibitem{wright15-1}
	L.~G. Wright, W.~H. Rinneinger, D.~N. Christodoulides, and F.~W. Wise,
	\enquote{Spatiotemporal dynamics of multimode optical solitons,}
	{\protect\JournalTitle{Optics Express}} \textbf{23}, 3492--3506 (2015).
	
	\bibitem{sunY22}
	Y.~Sun, M.~Zitelli, M.~Ferraro, \emph{et~al.}, \enquote{Multimode soliton
		collisions in graded-index optical fibers,} {\protect\JournalTitle{Optics
			Express}} \textbf{30}, 21710--21724 (2022).
	
	\bibitem{sunY24}
	Y.~Sun, P.~Parra-Rivas, G.~P. Agrawal, \emph{et~al.}, \enquote{Multimode
		solitons in optical fibers: a review,} {\protect\JournalTitle{Photonics
			Research}} \textbf{12}, 2581--2632 (2024).
	
	\bibitem{wright17}
	L.~G. Wright, D.~N. Christodoulides, and F.~W. Wise, \enquote{Spatiotemporal
		mode-locking in multimode fiber lasers,} {\protect\JournalTitle{Science}}
	\textbf{358}, 94--97 (2017).
	
	\bibitem{wright20}
	L.~G. Wright, P.~Sidorenko, H.~Pourbeyram, \emph{et~al.}, \enquote{Mechanisms
		of spatiotemporal mode-locking,} {\protect\JournalTitle{Nature Physics}}
	\textbf{16}, 565--–570 (2020).
	
	\bibitem{safaei20}
	R.~Safaei, G.~Fan, O.~Kwon, \emph{et~al.}, \enquote{High-energy
		multidimensional solitary states in hollow-core fibres,}
	{\protect\JournalTitle{Nature Photonics}} \textbf{14}, 733--739 (2020).
	
	\bibitem{piccoli21}
	R.~Piccoli, J.~M. Brown, Y.-G. Jeong, \emph{et~al.}, \enquote{Intense few-cycle
		visible pulses directly generated via nonlinear fibre mode mixing,}
	{\protect\JournalTitle{Nature Photonics}} \textbf{15}, 884--889 (2021).
	
	\bibitem{brahms22}
	C.~Brahms and J.~C. Travers, \enquote{Soliton self-compression and resonant
		dispersive wave emission in higher-order modes of a hollow capillary fibre,}
	{\protect\JournalTitle{Journal of Physics: Photonics}} \textbf{4}, 034002
	(2022).
	
	\bibitem{graini23}
	L.~Graini and B.~Ortac, \enquote{Spatial beam self-cleaning accompanied by
		self-similar propagation in few-mode graded-index fiber,}
	{\protect\JournalTitle{JOSA B}} \textbf{40}, 2139--2145 (2023).
	
	\bibitem{poletti08}
	F.~Poletti and P.~Horak, \enquote{Description of ultrashort pulse propagation
		in multimode optical fibers,} {\protect\JournalTitle{Journal of the Optical
			Society of America B}} \textbf{25}, 1645--1654 (2008).
	
	\bibitem{tani14}
	F.~Tani, J.~C. Travers, and P.~S. Russell, \enquote{Multimode ultrafast
		nonlinear optics in optical waveguides: numerical modeling and experiments in
		kagomé photonic-crystal fiber,} {\protect\JournalTitle{Journal of the
			Optical Society of America B}} \textbf{31}, 311--320 (2014).
	
	\bibitem{bejot19}
	P.~Béjot, \enquote{Multimodal unidirectional pulse propagation equation,}
	{\protect\JournalTitle{Physical Review E}} \textbf{99}, 032217 (2019).
	
	\bibitem{wright18}
	L.~G. Wright, Z.~M. Ziegler, P.~M. Lushnikov, \emph{et~al.}, \enquote{Multimode
		nonlinear fiber optics: Massively parallel numerical solver, tutorial, and
		outlook,} {\protect\JournalTitle{IEEE Journal of Selected Topics in Quantum
			Electronics}} \textbf{24}, 5100516 (2018).
	
	\bibitem{github_GMMNLSE}
	\enquote{{GMMNLSE},} \url{https://github.com/WiseLabAEP/GMMNLSE-Solver-FINAL}.
	
	\bibitem{ferraro23}
	M.~Ferraro, F.~Mangini, R.~Jauberteau, \emph{et~al.}, \enquote{Spatial beam
		self-cleaning in multimode fibers: the role of light polarization,}
	{\protect\JournalTitle{SPIE}} \textbf{12407} (2023).
	
	\bibitem{niuJ07}
	J.~Niu and J.~Xu, \enquote{Coupling efficiency of laser beam to multimode
		fiber,} {\protect\JournalTitle{Optics Communications}} \textbf{274},
	315–--319 (2007).
	
	\bibitem{guangZ18}
	Z.~Guang and Y.~Zhang, \enquote{Coupling ultrafast laser pulses into few-mode
		optical fibers: a numerical study of the spatiotemporal field coupling
		efficiency,} {\protect\JournalTitle{Applied Optics}} \textbf{57}, 9835--9844
	(2018).
	
	\bibitem{jolly23-2}
	S.~W. Jolly and P.~Kockaert, \enquote{Coupling to multi-mode waveguides with
		space-time shaped free-space pulses,} {\protect\JournalTitle{Journal of
			Optics}} \textbf{25}, 054002 (2023).
	
	\bibitem{ahsan20}
	A.~S. Ahsan and G.~P. Agrawal, \enquote{Effect of an input beam's shape and
		curvature on the nonlinear effects in graded-index fibers,}
	{\protect\JournalTitle{Journal of the Optical Society of America B}}
	\textbf{37}, 858--867 (2020).
	
	\bibitem{fabert20}
	M.~Fabert, M.~S{\u{a}}p{\^a}nțan, K.~Krupa, \emph{et~al.}, \enquote{Coherent
		combining of self-cleaned multimode beams,} {\protect\JournalTitle{Scientific
			reports}} \textbf{10}, 20481 (2020).
	
	\bibitem{akturk10}
	S.~Akturk, X.~Gu, P.~Bowlan, and R.~Trebino, \enquote{Spatio-temporal couplings
		in ultrashort laser pulses,} {\protect\JournalTitle{Journal of Optics}}
	\textbf{12}, 093001 (2010).
	
	\bibitem{gu04}
	X.~Gu, S.~Akturk, and R.~Trebino, \enquote{Spatial chirp in ultrafast optics,}
	{\protect\JournalTitle{Optics Communications}} \textbf{242}, 599--604 (2004).
	
	\bibitem{quere14}
	F.~Qu{\'e}r{\'e}, H.~Vincenti, A.~Borot, \emph{et~al.}, \enquote{Applications
		of ultrafast wavefront rotation in highly nonlinear optics,}
	{\protect\JournalTitle{Journal of Physics B: Atomic, Molecular and Optical
			Physics}} \textbf{47}, 124004 (2014).
	
\end{thebibliography}
\end{document}